# *Improved oxidation resistance of high emissivity coatings on fibrous ceramic for reusable space systems*

**Gaofeng Shao[a,b,c,*], Yucao Lu[a,b], Dorian A. H. Hanaor[c], Sheng Cui[a,b], Jian Jiao[d], Xiaodong Shen[a,b,*]**

*a. State Key Laboratory of Materials–Oriented Chemical Engineering, College of Materials Science and Engineering, Nanjing Tech University, Nanjing 210009, China*

*b. Jiangsu Collaborative Innovation Center for Advanced Inorganic Function Composites, Nanjing 210009, China*

*c. Fachgebiet Keramische Werkstoffe, Institut für Werkstoffwissenschaften und –technologien, Technische Universität Berlin, Berlin 10623, Germany*

*d. National Key Laboratory of Advanced Composite, Beijing Institute of Aeronautical Materials, Beijing 101300, China*

*Corresponding authors: Gaofeng Shao, Xiaodong Shen

Tel.: +86 25 83587234; fax: +86 25 83221690

E–mail addresses:

Gaofeng.shao@ceramics.tu-berlin.de (G. Shao),

xdshen@njtech.edu.cn (X. Shen)

**Abstract:**

To develop high emissivity coatings on fibrous ceramic substrates with improved thermal resistance for reusable space systems, $WSi_2$–$MoSi_2$–Si–$SiB_6$-borosilicate glass coatings were prepared on fibrous $ZrO_2$ by slurry dipping and subsequent high temperature rapid sintering. A coating with 20 wt% $WSi_2$ and 50 wt% $MoSi_2$ presents optimal thermal stability with only 10.06 mg/cm$^2$ mass loss and 4.0% emissivity decrease in the wavelength regime 1.27–1.73 μm after 50 h oxidation at 1773 K. The advantages of double phase metal-silicide coatings combining $WSi_2$ and $MoSi_2$ include improved thermal compatibility with the substrate and an enhanced glass-mediated self-healing ability.

## 1. Introduction

Fibrous ceramics are of growing interest in aerospace applications, particularly towards reusable thermal protection systems (RTPSs) in next generation hypersonic vehicles, due to their attributes of low density, low thermal conductivity and high thermal stability [1]. A typical concept for a RTPS structure comprises a high emissivity coating (HEC) on the surface of low thermal conductivity fibrous ceramic insulation, facilitating radiative cooling while minimizing heat conduction. In the past few years, fibrous $ZrO_2$ ceramics have been the subject of increasing attention towards thermal insulation applications in ultra–high temperature environments because of their higher temperature stability relative to fibrous quartz or mullite ceramics [2]. Consequently, HEC materials with similarly high thermal stability are sought to complement fibrous $ZrO_2$ insulation materials in RTPSs.

Among existing coating systems, single phase $MoSi_2$ based HECs have been applied successfully on the surfaces of various substrates including fibrous $SiO_2$, mullite and $ZrO_2$ ceramics, with favorable emissivity levels of 0.85 and higher achieved [3-10]. In our previous work, $MoSi_2$–borosilicate glass coating prepared on fibrous $ZrO_2$ ceramic showed excellent thermal shock behavior between room temperature and 1673 K [8], and following the inclusion of $ZrO_2$ in the coating material, this thermal shock resistance was extended to 1773 K [9]. However, the emissivity values of both coating types were only moderately higher than 0.8, and the coating surfaces were nearly fully oxidized following thermal shock tests, which had a negative effect on radiative performance. Recently, the inclusion of $TaSi_2$ in $MoSi_2$ based HEC systems was explored and was found to alter their radiative properties and thermal endurance [7, 11]. Specifically, in our previous research, a metal silicide–glass hybrid coating was prepared on fibrous $ZrO_2$ by a rapid sintering method [11]. The coating, comprising an outer Ta–Si–O glass layer and a dense $MoSi_2$–$TaSi_2$–glass inner layer was found to be thermally stable in air at 1773 K in air for 50 h, with a total emissivity around 0.9 in the 0.3–2.5 μm wavelength regime. Its thermal stability and consistent emissivity were attributed to the optimized viscosity of the Ta–Si–O layer and the self–healing ability of the $MoSi_2$–$TaSi_2$–glass. To date, most investigations on high emissivity transition-metal disilicides have focused on $MoSi_2$ and $TaSi_2$, while $WSi_2$ has received relatively limited attention. The interactions of $MoSi_2$ and $WSi_2$ have not been reported in terms of emissive performance and the thermal stability of radiative properties has not been adequately studied in such systems. This is particularly important as ablation and oxidation alter the surface composition and microstructure of coatings, which may affect their radiative performance [12, 13]. Towards aerospace applications, HECs are required with intrinsic high temperature oxidation resistance, which limits the attenuation of coating emissivity during long–term exposure to high temperature oxidative atmospheres.

In general, defect formation and defect healing occur competitively during the oxidation process [14, 15]. Owing to the formation of a flowing glass phase at high temperatures, defects are healed to a certain extent, thus limiting the oxidation driven degradation of emissive phases and preserving performance. The self-healing ability of glass phase is determined not only by its relative quantity, but also its viscosity. For this reason viscosity is one of the most important properties for the glass phase generated on the coating surface during high temperature oxidation[16]. Amorphous $SiO_2$ is known to exist as a dense and highly viscous glass, which is strongly networked due to the sharing of oxygen atoms between neighboring $SiO_4$ tetrahedra. Boron, added as $B_2O_3$, acts as a network modifier in $SiO_2$ and reduces its viscosity by the production of non-bridging oxygen atoms (NBO). Accordingly, borosilicate glass has emerged as a promising oxidation resistant sealant material, due to its low oxygen permeability, appropriate viscosity and fluidity [17, 18], which plays a critical role in sealing defects of coatings towards oxidation.

In this work, in order to improve thermal resistance of HECs on fibrous $ZrO_2$, $MoSi_2$ and $WSi_2$ were simultaneously introduced as double phase emissive agents. $WSi_2$–$MoSi_2$–Si –$SiB_6$–borosilicate glass multiphase coatings with different compositions were prepared by slurry dipping and subsequent high temperature rapid sintering. Specimens were exposed to a high temperature oxidizing atmosphere for 50 h. Microstructure, phase composition and radiative properties, before and after oxidation, were investigated and the oxidation mechanism was examined.

## 2. Experimental

### 2.1 Preparation of the coatings

Fibrous $ZrO_2$ ceramics with dimensions of 25 mm×25 mm×5 mm (Anhui Crystal New Materials Co. Ltd., China) were used as substrates. The coatings were prepared by slurry dipping and rapid sintering. Firstly, borosilicate glass (BSG) was prepared by high temperature melting and water quenching, as reported in our previous work [8]. Then, a slurry was prepared by ball–milling as follows: $MoSi_2$, $WSi_2$, BSG, Si, $SiB_6$, ethanol and a 1 wt% aqueous solution of carboxyl methylcellulose (CMC) were mixed in a nylon ball–mill and milled for 6 h at a rotation speed of 400 rpm. The mass ratio of the powders, ethanol, CMC aqueous solution and zirconia balls was 1.4:1:0.25:2. Coatings were prepared on the top surfaces and four sides of treated fibrous $ZrO_2$ substrates by slurry dipping: the top surface and two opposing sides were dip–coated first, and then the coated sample was dried at 333 K for 30 min and removed to immerse the other two sides. The coating thickness was controlled by repeating such dipping four times, and the dwell time for each dipping was 2 s. Following these dipping cycles, the coated samples were dried at 333 K for 12 h and then at 373 K for 6 h. Finally, the fully dried samples were fired in a pre-heated furnace at 1773 K. The samples were held for 15 min in natural air and later removed from the furnace for rapid cooling. The compositions and sample names of the powders are shown in Table 1. Samples are referred to in terms of their respective weight percentages of $WSi_2$ and/or $MoSi_2$.

**Table 1. Compositions and labels of the coatings studied in this work**

| Samples | Composition (Wt. %) | | | | |
|---|---|---|---|---|---|
| | **$MoSi_2$** | **$WSi_2$** | **Borosilicate glass** | **Si** | **$SiB_6$** |
| **W70** | 0 | 70 | 18.5 | 10 | 1.5 |
| **M20W50** | 20 | 50 | 18.5 | 10 | 1.5 |
| **M35W35** | 35 | 35 | 18.5 | 10 | 1.5 |
| **M50W20** | 50 | 20 | 18.5 | 10 | 1.5 |
| **M70** | 70 | 0 | 18.5 | 10 | 1.5 |

**2.2 Isothermal oxidation**

The high temperature oxidation resistance of coated fibrous $ZrO_2$ specimens were investigated by a static isothermal oxidation test, which was carried out in ambient air at 1773 K. Coated specimens were loaded on a $ZrO_2$ refractory brick and transferred into a furnace that had been pre-heated to 1773 K. Specimens were then taken out at designated times (1 h, 5 h, 10 h, 20 h and 50 h) and cooled to room temperature. Coated specimens were weighed with ±0.1 mg accuracy and returned to the furnace for further isothermal oxidation tests. The mass change per unit area at different oxidation times were calculated according to eq. (1) and averaged over three specimens.

$$\varDelta W = \frac{m_0 - m_1}{A} \quad (1)$$

Here, $\varDelta W$ is the mass change per unit area; $m_0$ and $m_1$ are the weights of the specimens before and after oxidation, respectively; and $A$ is the surface area of the coated sample.

**2.3 Radiative properties**

Reflectance spectra in the wavelength range of 0.3–2.5 μm were obtained by ultraviolet–visible–near infrared spectrophotometer (UV−Vis–NIR, Cary 5000, Varian, CA) equipped with a $BaSO_4$ integrating sphere. The reflectance spectra in the wavelength range of 2.5–15 μm were by Fourier transform infrared spectroscopy (FT–IR, Frontier, PerkinElmer LLC) equipped with a gold–coated integrating sphere. The emissivity is obtained indirectly from the measured reflectance based on the relationship of materials, ε= A= 1–R–T, where ε, A, R and T are the emissivity, absorptivity, reflectivity, and transmissivity respectively. The total emissivity can be derived from the reflectance spectrum according to Eq. (2)

$$\varepsilon_{\mathrm{T}} = \frac{\int_{\lambda_1}^{\lambda_2} \left[1 - R(\lambda)\right] P_{\mathrm{B}}(\lambda) \mathrm{d}\lambda}{\int_{\lambda_1}^{\lambda_2} P_{\mathrm{B}}(\lambda) \mathrm{d}\lambda} \quad (2)$$

Where $\lambda$ is the wavelength, $R(\lambda)$ is the reflectance, $P_B(\lambda)$ is given by Planck's law according to Eq. (3):

$$P_{\mathrm{B}}(\lambda) = \frac{C_1}{\lambda^5 \left[\exp\left(\frac{C_2}{\lambda T}\right) - 1\right]} \quad (3)$$

Where $C_1$=3.743×10$^{-16}$Wm$^2$, $C_2$=1.4387×10$^{-2}$mK. The calculation of the total emissivity was performed in a MATLAB environment using Eq. (2) and (3).

### 2.4 Characterization

Phase compositions of the coating surface were identified using a Rigaku Miniflex X–ray diffractometer (XRD) with Cu–Kα radiation (λ= 0.15406 nm). The microstructure and the elemental distribution were characterized and analyzed by a scanning electron microscope (SEM, JSM–6510, JEOL, Japan) equipped with energy dispersive X-ray spectroscopy (EDS). The surface morphology was observed using a confocal laser scanning microscope (CLSM, Olympus LEXT OLS 4000, Japan). The CLSM images were obtained on a microscope powered by a singer laser (λ = 405 nm) in the reflected light mode. X-ray photoelectron spectroscopy (XPS) spectra were obtained using a Theta Probe Thermo Fisher Scientific spectrometer with a monochromatic Al Ka X-ray source (1486.6 eV).

## 3. Results

### 3.1 Phase compositions and microstructures of the coatings

XRD patterns of the as–prepared coatings are shown in Fig.1. The coatings are composed of amorphous glass phase alongside tungsten or molybdenum silicides, $WSi_2$, $MoSi_2$, $W_5Si_3$ and $Mo_5Si_3$, cristobalite $SiO_2$, metallic W and Mo and small amounts of elemental Si. Metallic W or Mo and $W_5Si_3$ and $Mo_5Si_3$ phases are the oxidation products of disilicides following reactions (4) and (5) [19].

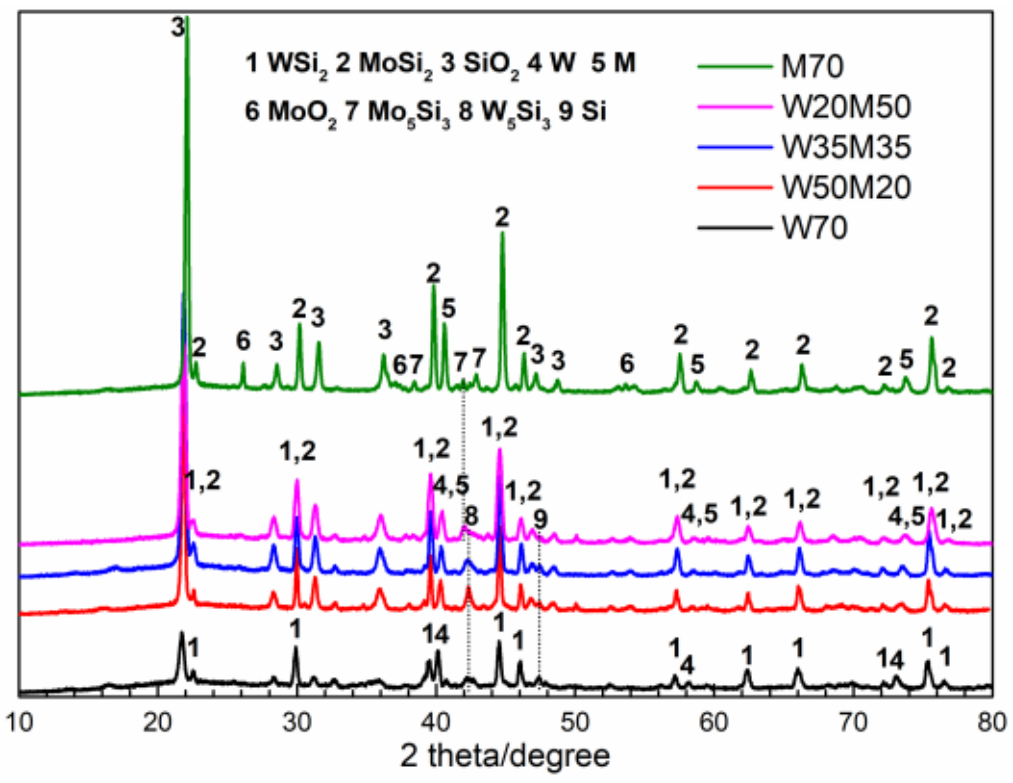

**Fig.1 XRD patterns of the as–prepared coatings**

The $MoO_2$ phase appears in the M70 coating, through the oxidation of $MoSi_2$ following reaction (6). However, $WO_2$ is not found in the W70 coatings due to its narrow stability domain [20]. During the rapid sintering process, as disilicide particles are contained within the borosilicate glass, an environment with low oxygen content is provided for the oxidation of $WSi_2$ and $MoSi_2$. Therefore, W, Mo, $W_5Si_3$ and $Mo_5Si_3$ phases are preferentially formed rather than $WO_3$ or $MoO_3$ following reaction (7) [19, 21], which can effectively suppress the oxidation of $WSi_2$ and $MoSi_2$ with the evaporation of $WO_3$ or $MoO_3$. Cristobalite is formed not only from the oxidation of $WSi_2$ and $MoSi_2$ according to Eqs. (4) – (8), but also from the partial borosilicate glass crystallized during cooling, especially in coatings with higher $MoSi_2$ content. The corresponding Gibbs free energy ($\Delta G^0$) under standard conditions of the reactions (4)-(8) at temperatures ranging from 373 K to 1773 K were calculated using the Reaction Web module of FactSage software. As shown in Fig. 2, the $\Delta G^0$ values for all reactions are negative, indicating that the reactions can proceed spontaneously. This parameter helps to understand formation process of the as–prepared coatings.

5/7 $(W/Mo)Si_2(s)$ + $O_2(g)$ = 1/7 $(W/Mo)_5Si_3(s)$ + $SiO_2(s)$ (4)

1/2 $(W/Mo)Si_2(s)$ + $O_2(g)$ = 1/2 W/Mo (s) + $SiO_2(s)$ (5)

1/3 (W/Mo) $Si_2(s)$ + $O_2(g)$ = 1/3 $(W/Mo)O_2(s)$ + 2/3 $SiO_2(s)$ (6)

2/7 $(W/Mo)Si_2(s)$ + $O_2(g)$ = 2/7 $(W/Mo)O_3$ (g) + 4/7 $SiO_2(s)$(7)

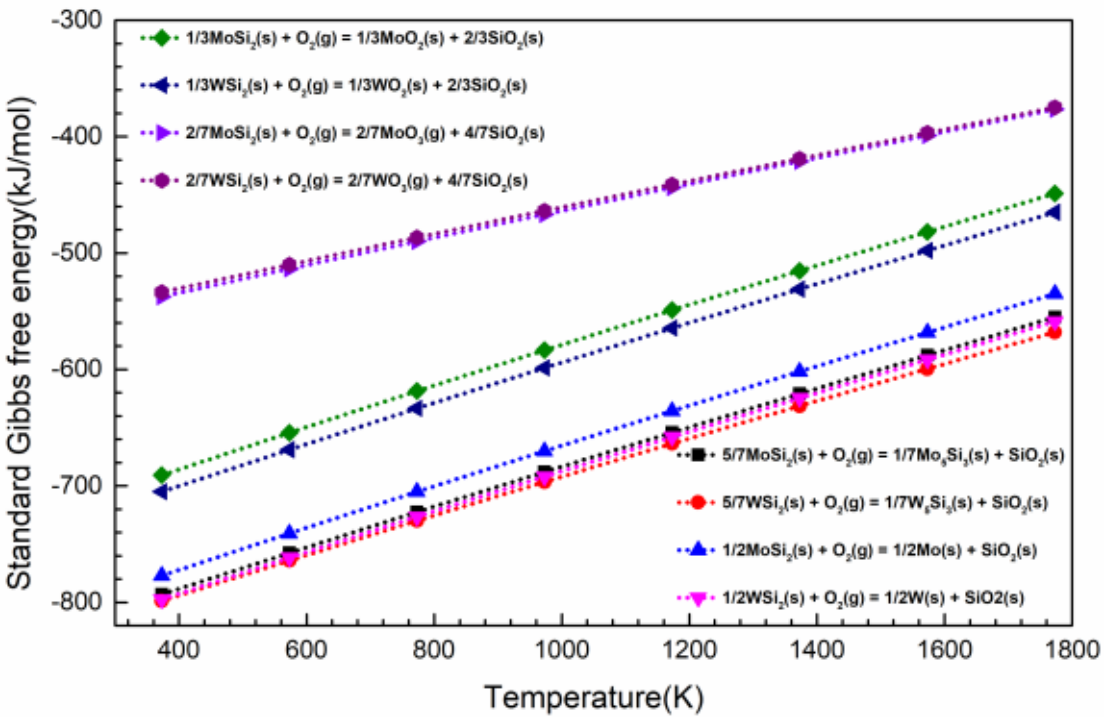

**Fig.2 Changes of standard Gibbs free energies of reactions (4)–(7) at different temperatures calculated by the Reaction Web module of FactSage.**

Cross–sectional SEM micrographs of the as–prepared coating (W20M50), along with the fibrous substrate are shown in Fig. 3. The overall coating thickness is found to be in the order 80-100μm. Three distinctive microstructures observed in the coating are (A) the top layer, (B) the main layer and (C) the interfacial transition layer, which are further magnified in Fig. 3(b–d).

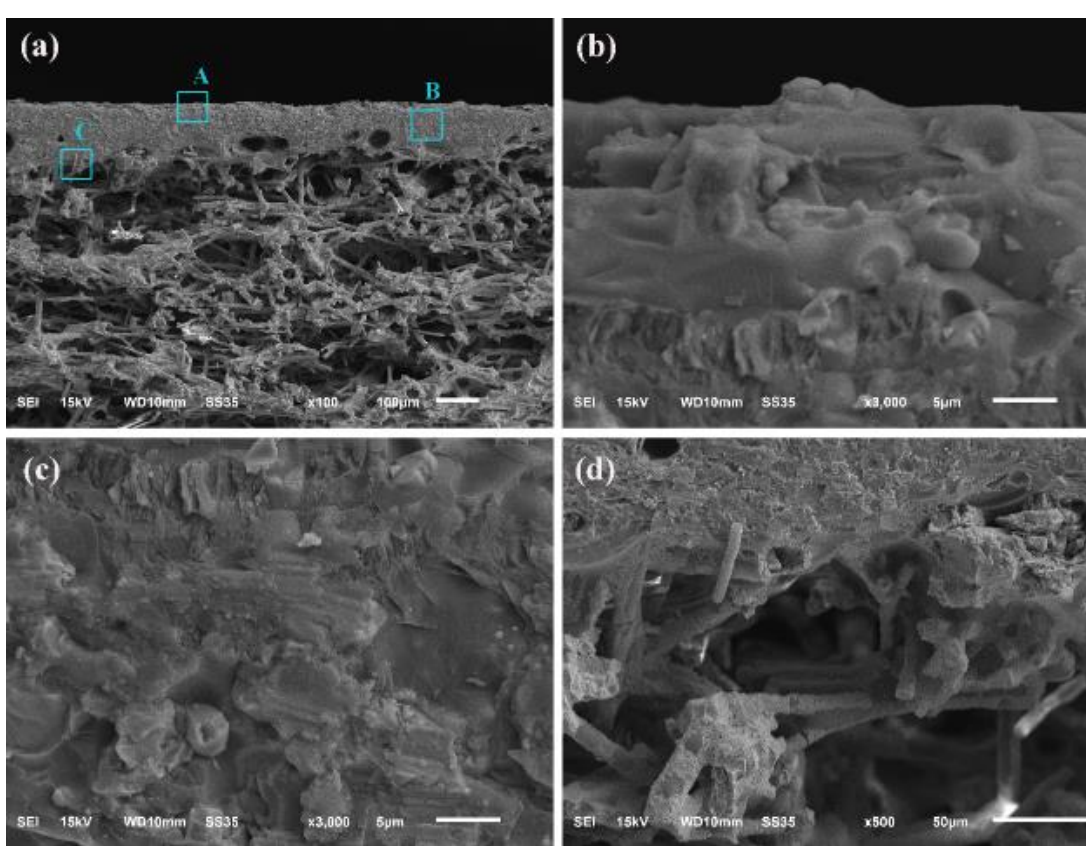

**Fig.3 (a) Cross–sectional SEM images of the as–prepared coating (W20M50); (b), (c), (d) The magnified SEM images of region A, B and C in (a).**

A BSE SEM micrograph of a polished cross-section embedded in resin shows the morphology of the as–prepared W20M50 coating in Fig. 4(a), revealing three phases (in addition to the upper and lower resin layers), distinguished by their greyscale levels. Elemental mapping (Fig. 4 (b)–(h)) shows that $MoSi_2$ and $WSi_2$ are surrounded by the borosilicate glass.

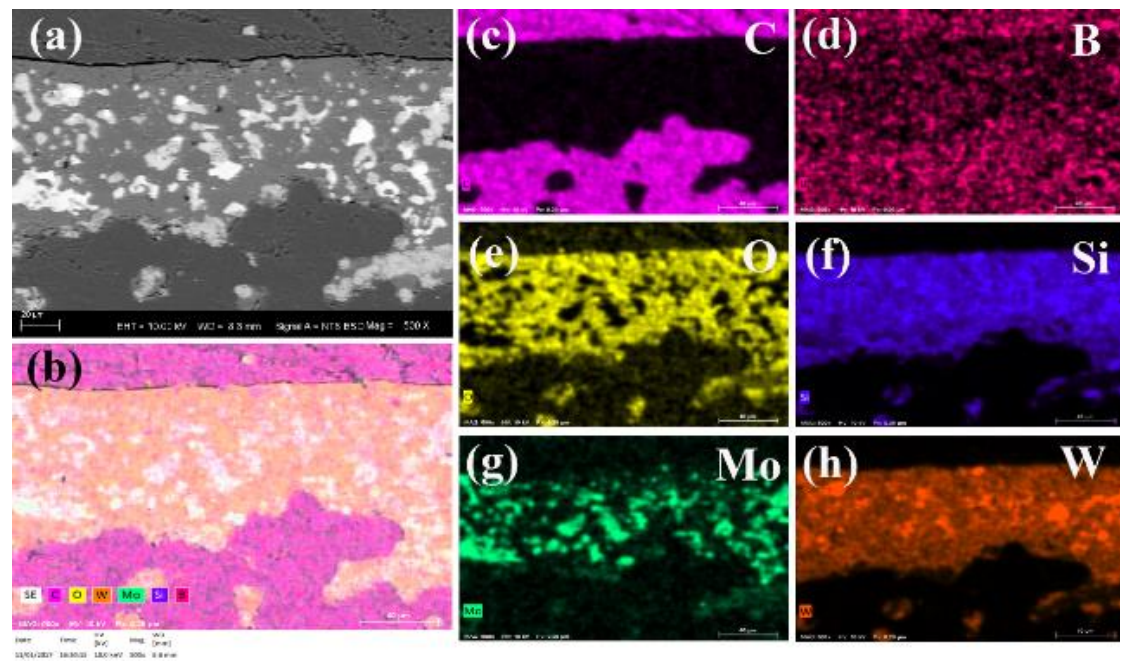

**Fig.4 (a) BSE image and (b–h) EDS mappings of the polished cross–section of W20M50**

### 3.2 Radiative properties of coatings

In order to investigate the radiative properties of the $WSi_2$–$MoSi_2$ based coatings after long-term high-temperature oxidation, an isothermal oxidation test at 1773 K in air was carried out. Fig. 5 compares the spectral emissivity curves of the coatings before and after oxidation at 1773 K for 50 h in the wavelength range of 0.3–2.5 μm and 2.5–15 μm.

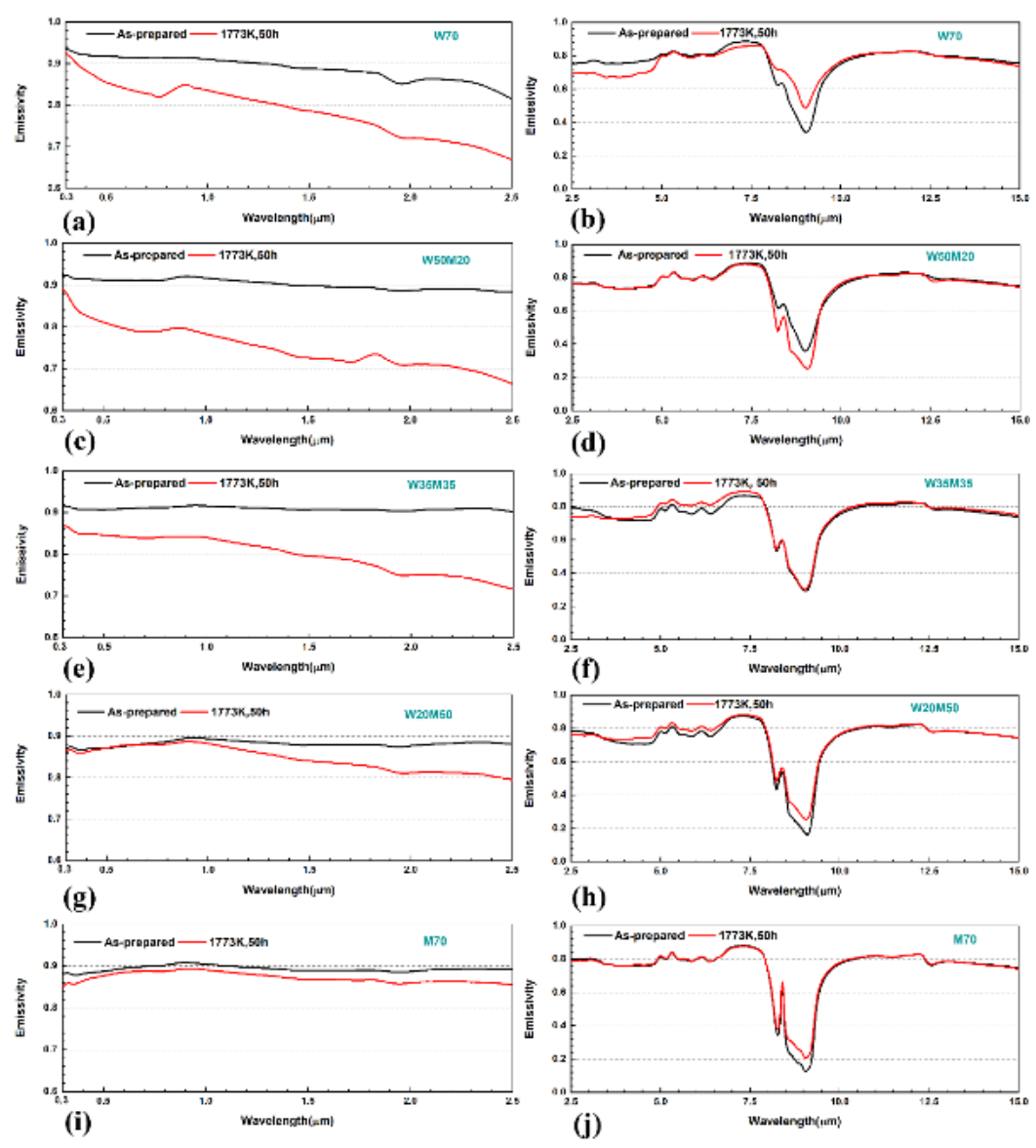

**Fig.5 Spectral emissivity curves of coatings before and after oxidation at 1773 K for 50 h in the wavelength range of 0.3–2.5 μm and 2.5–15 μm**

In the wavelength regime 0.3–2.5 μm the spectral emissivity curves of all as–prepared coatings with different compositions present similar features, and emissivity values are above 0.85. Over the wavelength range of 2.5–15 μm, emissivity varies in the region of 0.8 with a moderate shoulder around 9 μm, corresponding to the asymmetrical Si–O stretching vibrations of $[SiO_4]$ tetrahedra. The phenomenon is more prominent the amorphous $SiO_2$–based glass phase rather than the silicide [22-24]. Fig. 6 shows the histogram of the calculated total emissivity at various wavelengths according to Eqs. (1) and (2). For wavelengths in the range 0.3–2.5 μm, the total emissivity values of the five as–prepared coatings with increasing $MoSi_2$ content are 0.925, 0.915, 0.910, 0.869 and 0.881. In the wavelength range of 3–5 μm, the emissivity values are greater than 0.7. In contrast, in the range of 8–14 μm, the emissivity is only ~0.6. After oxidation at 1773 K for 50 h, emissivity values of the different $WSi_2$–$MoSi_2$ based coatings are reduced to varying degrees.

Planck's radiation law given in Eq. (8), shows the spectral emissive power as a function of wavelength and temperature.

$$E_{\lambda b}(T) = \frac{2\pi C_1}{n^2 \lambda^5 \left[ \exp\left(\frac{C_2}{n\lambda T}\right) - 1 \right]} \quad (8)$$

This relationship shows that in general total radiated energy increases at higher temperatures and the intensity peak of the emitted spectrum shifts to shorter wavelengths. Additionally, energy emitted at shorter wavelengths increases more rapidly with temperature than the energy at longer wavelengths. As an example, for an ideal blackbody at 1273 K, 97 % of the energy emitted in thermal equilibrium is below 14 μm, and 76 % is below 5 μm[25].

As per Wien's displacement law (9):

$$\lambda_{max} = \frac{b}{T}, b = 2.898 \times 10^{-3} mK \quad (9)$$

This equation shows that in general, when the temperature increases from 1673 K to 2273 K, the largest radiated wavelength shifts from approximately 1.73 μm to 1.27 μm. For W70, W50M20 and W35M35 coatings, the total emissivity values decrease from 0.900, 0.900 and 0.909 to 0.792, 0.735 and 0.803, respectively, over the wavelength range 1.27–1.73 μm (histogram in red).

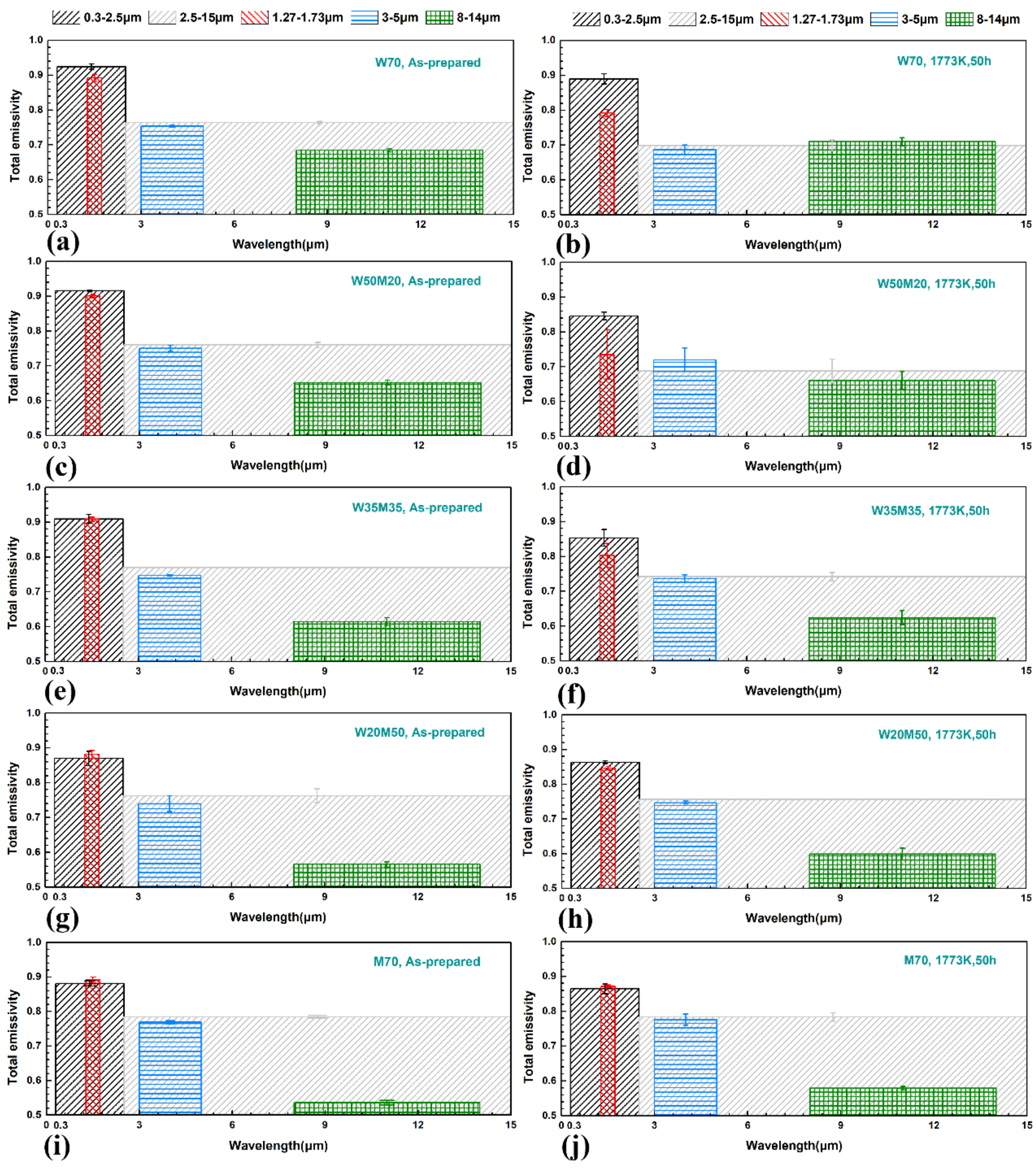

**Fig.6 Histograms of the calculated total emissivity in various wavelengths of the as-prepared coatings before and after oxidation at 1773 K for 50 h**

However, in the case of W20M50 and M70 coatings, the total emissivity values remain as high as 0.846 and 0.872 with only 4.0% and 2.1% decreases in emissivity, respectively. The attenuation of the coatings in 2.5–15 μm (histogram in grey) gradually decreases with the increase of $MoSi_2$ amount. The radiative loss of W70 coating in this wavelength region reaches to 8.6%, which is much higher than that of M70 coating (0.3%).

### 3.3 Oxidation resistance

#### 3.3.1 Isothermal oxidation curves

Fig.7 shows the isothermal oxidation curves of the coated samples at 1773 K for 50 h. It can be seen that the oxidation of the samples occurs as a continuous mass loss process. The process can be divided into two stages: the initial fast oxidation stage (0-10 h) and the slow stable oxidation stage (10-50 h). Neither a linear nor a parabolic rate law can be fitted. The mass loss per unit area of W70, W50M20, W35M35, W20M50 and M70 coatings are 6.70mg/cm$^2$, 7.79 mg/cm$^2$, 11.92 mg/cm$^2$, 10.06 mg/cm$^2$ and 14.96 mg/cm$^2$, respectively. Though the mass loss values of W70 and W50M20 coatings were the lowest, these coatings are not necessarily endowed with the best thermal resistance.

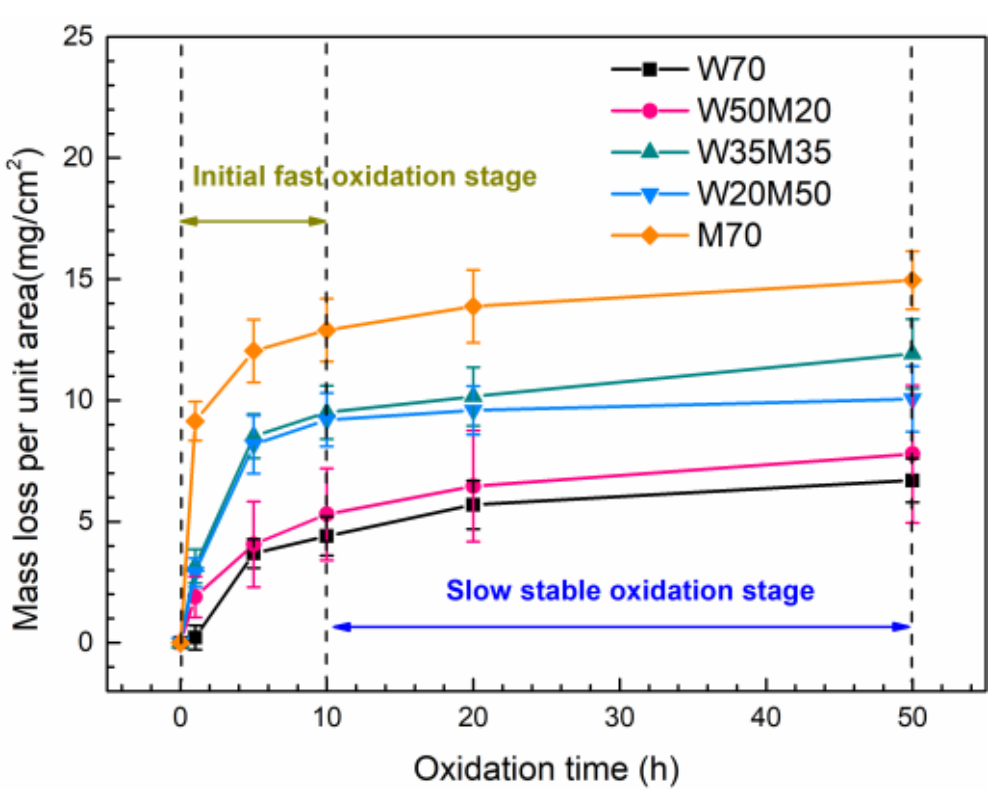

**Fig.7 Mass loss per unit area of coated samples during isothermal oxidation at 1773 K in air.**

#### 3.3.2 Phase compositions and chemical states of oxidized coatings

To examine the phase compositions of surfaces, the XRD patterns of the coated samples after oxidation at 1773 K for 50 h are shown in Fig. 8. Compared with the original coatings, the intensity of $WSi_2$ and $MoSi_2$ diffraction peaks decrease, and W, Mo, $W_5Si_3$ and $Mo_5Si_3$ diffraction peaks become stronger for all coatings.

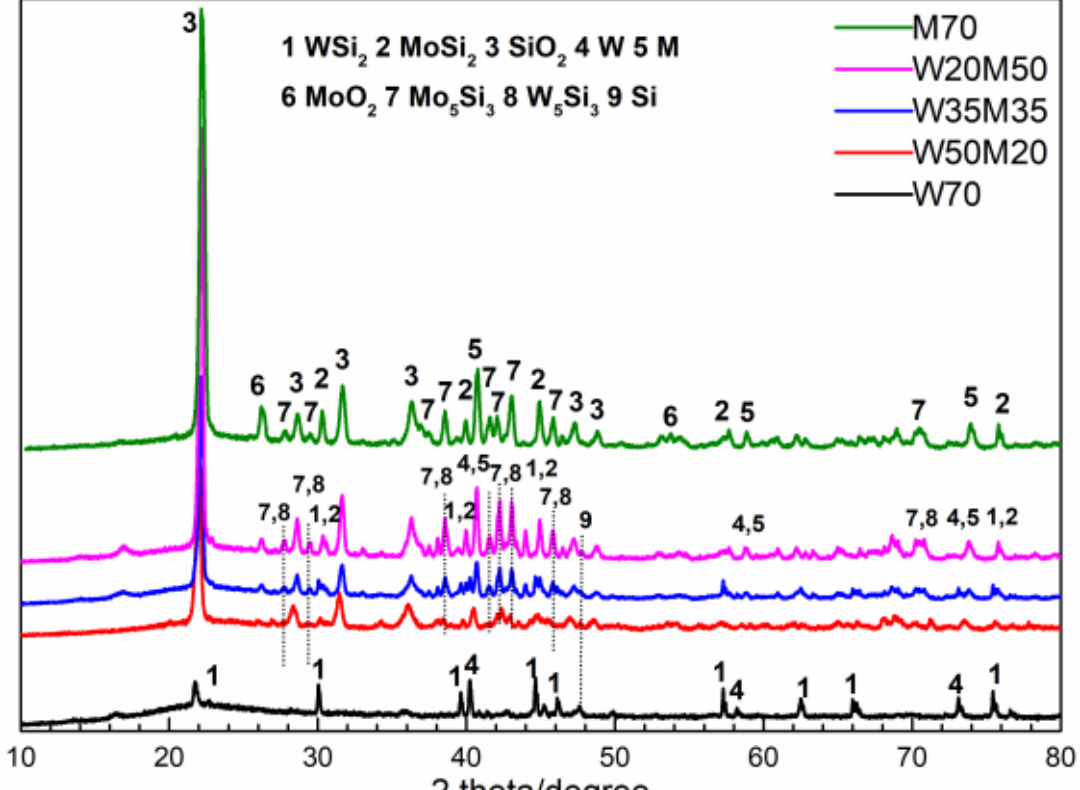

**Fig.8 XRD patterns of the different coatings after oxidation at 1773K for 50 h**

During the initial fast oxidation stage, the simultaneous oxidation of W, Mo and Si takes place according to reactions (7) and (12) [26]. The evaporation of $WO_3$ and $MoO_3$ leads to the rapid mass loss of the coatings. With prolonged oxidation time, a $SiO_2$–based glass compound scale is gradually formed on the outermost layer. This protective layer limits oxygen diffusion and inhibits the further oxidation of the inner coating, resulting in a flattening of the oxidation rate in the slow stable oxidation stage. In an environment with low oxygen pressure only Si can be oxidized as given in reactions (5) and (10) [27], and the oxidation of W and Mo could not take place when the coating was entirely covered by dense $SiO_2$-based glass layer [28]. According to the calculated standard Gibbs free energies of reactions (10) - (12) in Fig. 9, reactions (11) and (12) is preferred in an excessive oxygen environment, while the preferential oxidation of Si occurs in an oxygen-lean environment as indicated by reaction (10). It is the reason why W and Mo phases with strong intensity are detected in the coatings.

$1/3\ (W/Mo)_5Si_3(s)+O_2(g) = 5/3\ W/Mo\ (s)+SiO_2\ (s)$ (10)

$2/21\ (W/Mo)_5Si_3(s)+O_2(g) = 10/21\ (W/Mo)\ O_3\ (g)+6/21\ SiO_2\ (s)$ (11)

$2/3\ (W/Mo) + O_2(g) = 2/3\ (W/Mo)\ O_3\ (g)$ (12)

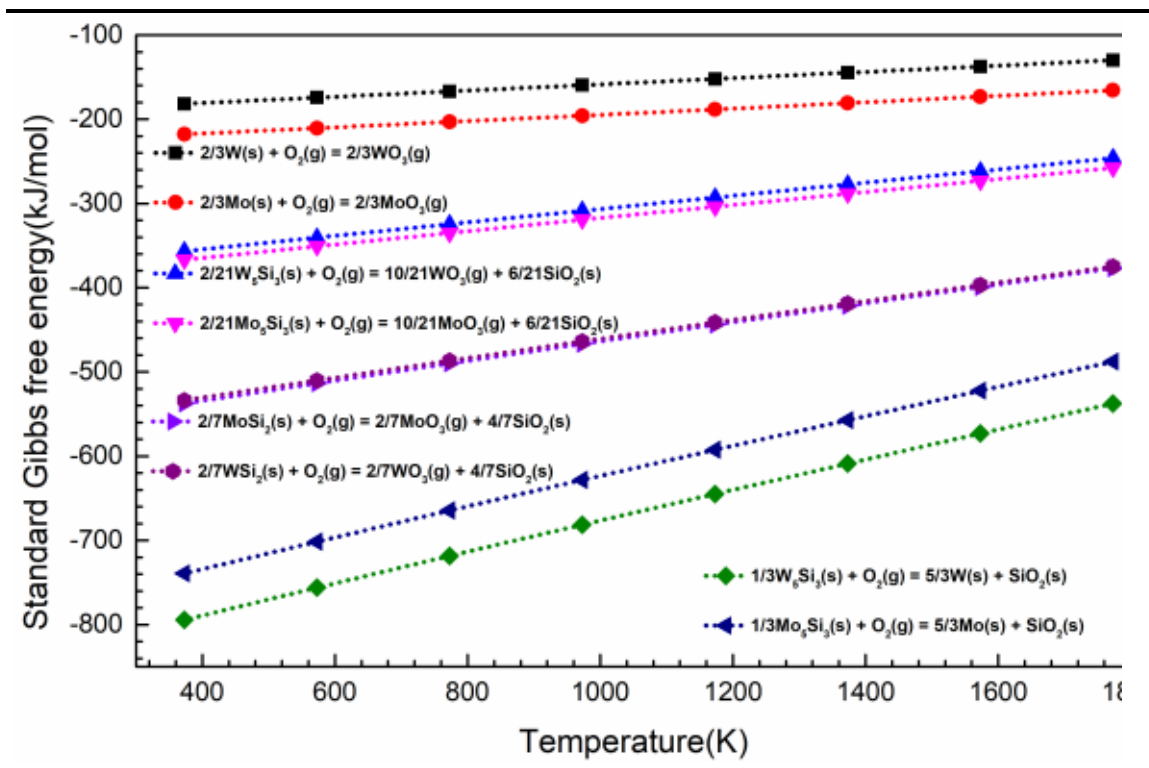

**Fig. 9 Changes of standard Gibbs free energies of reactions (7)–(10) at different temperatures calculated by the Reaction Web module of FactSage**

The chemical states of the coatings after oxidation are further investigated by XPS, this is particularly useful to interpret changes in the viscosity of the surface glass layer. Fig. 10 shows the high-resolution XPS W4f and Mo3d spectra of the coatings after oxidation. The W4f spectrum was analyzed in W70 and W20M50 coatings, and can be deconvoluted into one doublet. The binding energies of $W4f_{5/2}$ and $W4f_{7/2}$ are determined to be 37.4 eV and 35.2 eV in Fig.10 (a) and 37.8 eV and 35.8 eV in Fig. 10 (b), respectively, which is characteristic of $WO_3$[29]. However, no XPS signals of metallic W with the binding energies of $W4f_{5/2}$ (33.2 eV) and $W4f_{7/2}$ (31.0 eV) are detected.

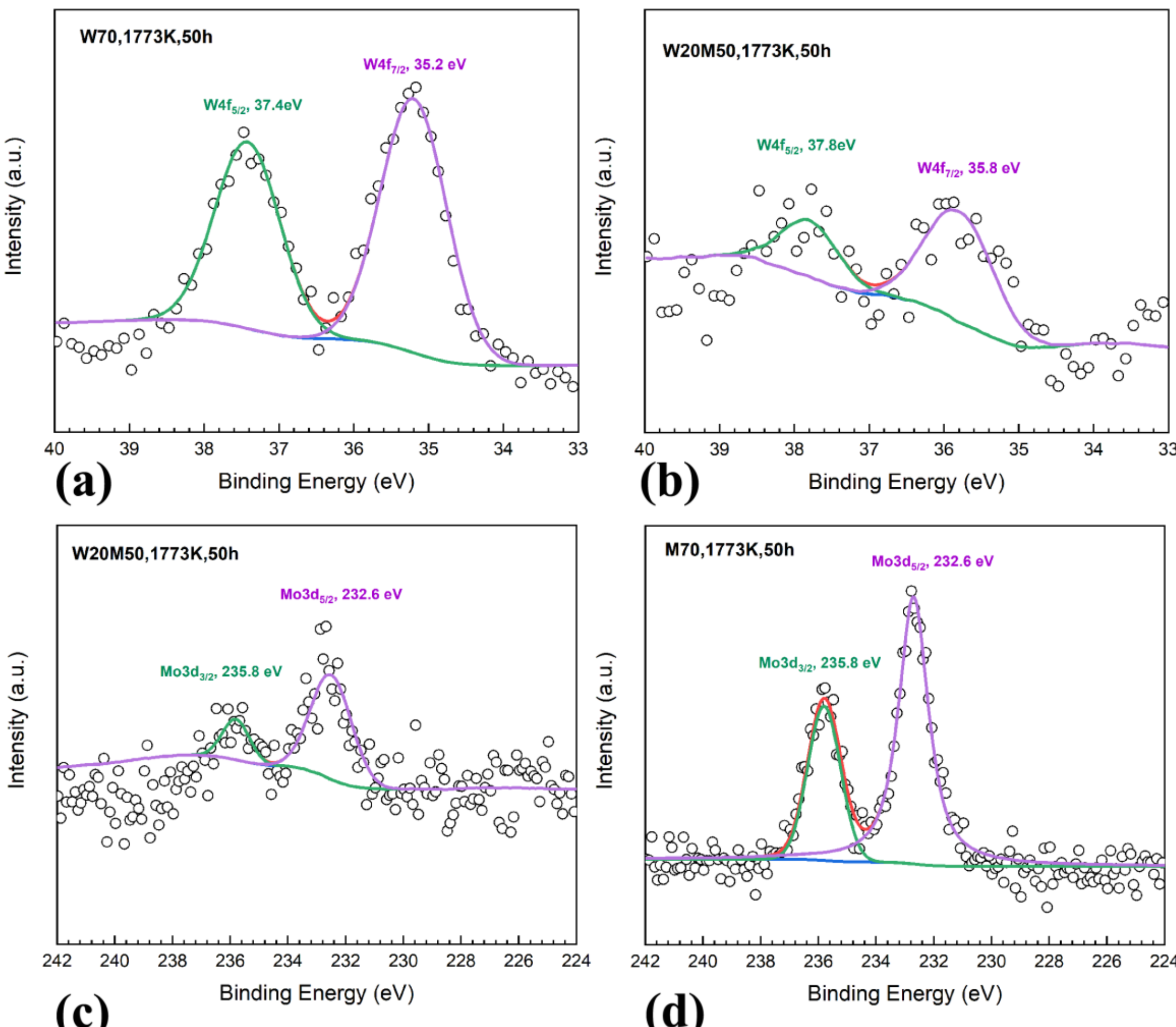

**Fig.10 XPS W4f and Mo3d spectra of the coatings after oxidation at 1773 K for 50 h, (a) W70; (b–c) W20M50; (d) M70**

Fig. 10 (c) and (d) show the high-resolution XPS Mo3d spectra of W20M50 and M70 coatings, and the binding energies of $Mo3d_{3/2}$ and $Mo3d_{5/2}$ are determined to be 235.8 eV and 232.6 eV, respectively, consistent with the presence of $MoO_3$[29, 30]. No $Mo3d_{5/2}$ values of 229.4 eV and 228 eV are obtained, implying the absence of Mo and $MoO_2$ phases on the surface of the coatings. However, W, Mo and $MoO_2$ phases were indeed detected by XRD (Fig. 8). This may be the result of the greater X-ray penetration depth, which is generally within tens of micrometers, while the photoelectrons escape depth of the coated sample is approximately a few nanometers. These results suggest that trace amounts of $WO_3$ and $MoO_3$ form at surfaces within depths of a few nanometers according to Eq. (10)–(11), while W, Mo and $MoO_2$ are generated in the coatings at depths in the order of tens of micrometers in accordance with Eq. (5), (6) and (10).

Viscosity is one the most important properties for the glass phase generated on the coating surface after high temperature oxidation. To date, only limited studies have been implemented to measure the viscosity of coatings. Previous research confirmed the effectiveness of XPS for directly measuring the relative concentrations of bridging oxygen (BO) and non–bridging oxygen (NBO) in the glass phase [11]. A bridging oxygen atom contains two bonds with energy near 532 eV, while non–bridging oxygen atoms occupy a single bond and a partial negative charge resulting in a shift in bond energy to approximately 530 eV [31]. Fig.11 shows the high-resolution XPS O1s spectra of the W70, W20M50 and M50 coatings after oxidation at 1773 K for 50 h. The binding energy values of BO are determined to be 532.2 eV, 532.3 eV and 532.3 eV, respectively. The NBO contribution of the three coatings is identified at the binding energy of 530.6 eV. Furthermore, the relative NBO concentration for these three coatings, corresponding to the area of the NBO peak over the total measured peak area, is 13.67%, 9.06% and 7.47%, respectively. The decrease in NBO concentration with increasing $MoSi_2$ content indicates that the glass viscosity of the coating increases. Based on XRD results, the cristobalite content increases with $MoSi_2$ content. It can be inferred that the crystalline $SiO_2$ incorporated in the oxide scale increases the viscosity of glass layer [32].

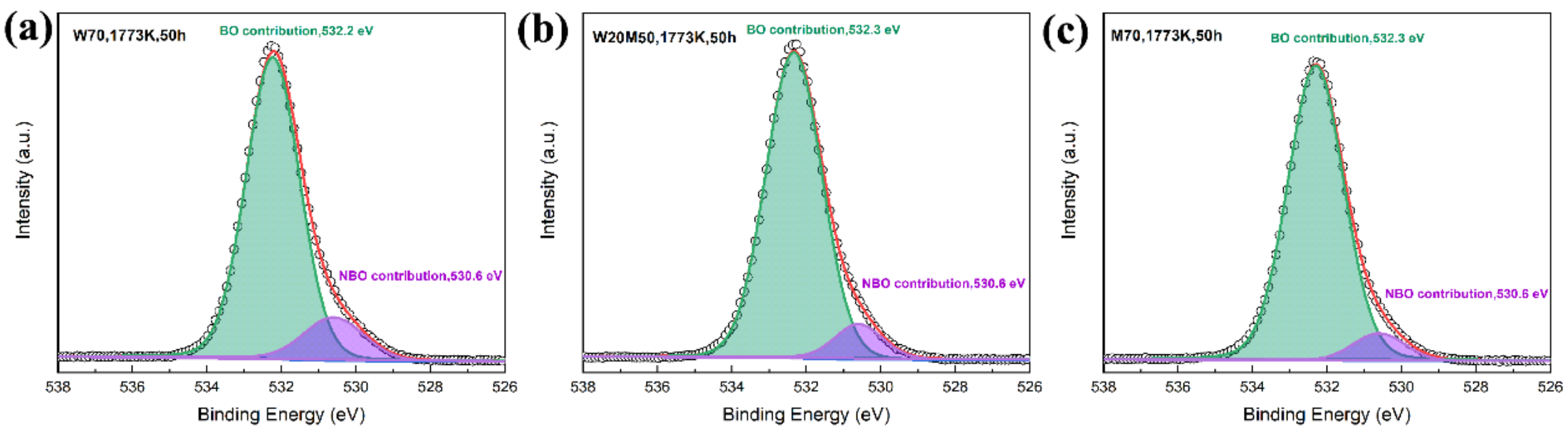

**Fig.11 High-resolution XPS O1s spectra of the coatings after oxidation at 1773 K for 50 h, (a) W70; (b) W20M50; (c) M70**

### 3.3.3 Microstructure following oxidation

Fig. 12 (a–j) show macroscopic images of the entire coated samples, as prepared, as well as following oxidation at 1773 K for 50 h. As shown in Fig.12 (a–e), the obtained coatings are dense, smooth and without cracks. However, the coatings exhibit different degrees of damage after long–term oxidation. Several significant cracks are generated on the surface of the W70 coating (Fig.12 (f)). With increasing $MoSi_2$ content, the size and amount of the cracks formed decrease, as seen in W50M20 and W35M35 (Fig. 12(g-h)). It is worth noting that no apparent cracks and no yellow phase are observed on the surface of W20M50 (Fig. 12(i)), which suggests that this material retains its structural integrity during oxidation. Relative to the surface of M70 (Fig. 12(j)), the surface of W20M50 is smoother and denser, implying that this sample exhibited optimal glass viscosity in the top layer, which plays a key role in suppressing and sealing the crack itself.

The surface and cross–section CLSM images of the coatings after oxidation are displayed in Fig. 12(k–t). As seen clearly in Fig.12 (p–q), the failure mode of W70 and W50M20 coatings includes delamination between the coating and the substrate, as the result of differential thermal expansion. When thermally induced stresses exceed the cohesive strength between the coating and the fibrous $ZrO_2$ layers, cracks and delamination occur in the interface. The failure of W70 and W50M20 coatings during high temperature oxidation can be predominantly attributed to the formation of cracks on the coating surface and delamination between the coating and the substrate. Delamination arises not only as the result of differential thermal expansion between the $WSi_2$–Si–$SiB_6$-borosilicate glass coating and the fibrous $ZrO_2$ substrate, but also from the volumetric swelling incurred by the formation of oxidation products (glass phase, $WO_3$, W, $W_5Si_3$) [14].

Surface SEM images of W70, W20M50 and M70 coatings after oxidation at 1773 K for 50 h are shown in Fig.13. In M70, further to surface macro-cracks (Fig. 12(f)) and delamination from the substrate (Fig. 12(p)), micro pores are observed within the glassy phase (Fig. 13(a) and (d)) as the result of gaseous oxides such as $B_2O_3$ and $MoO_3$ escaping from the low viscosity coating surface. As shown in Fig. 13(b) and (e), rather than pores, several nano-cracks are formed on the W20M50 coating surface, resulting from the volume expansion of the top oxidized layer.

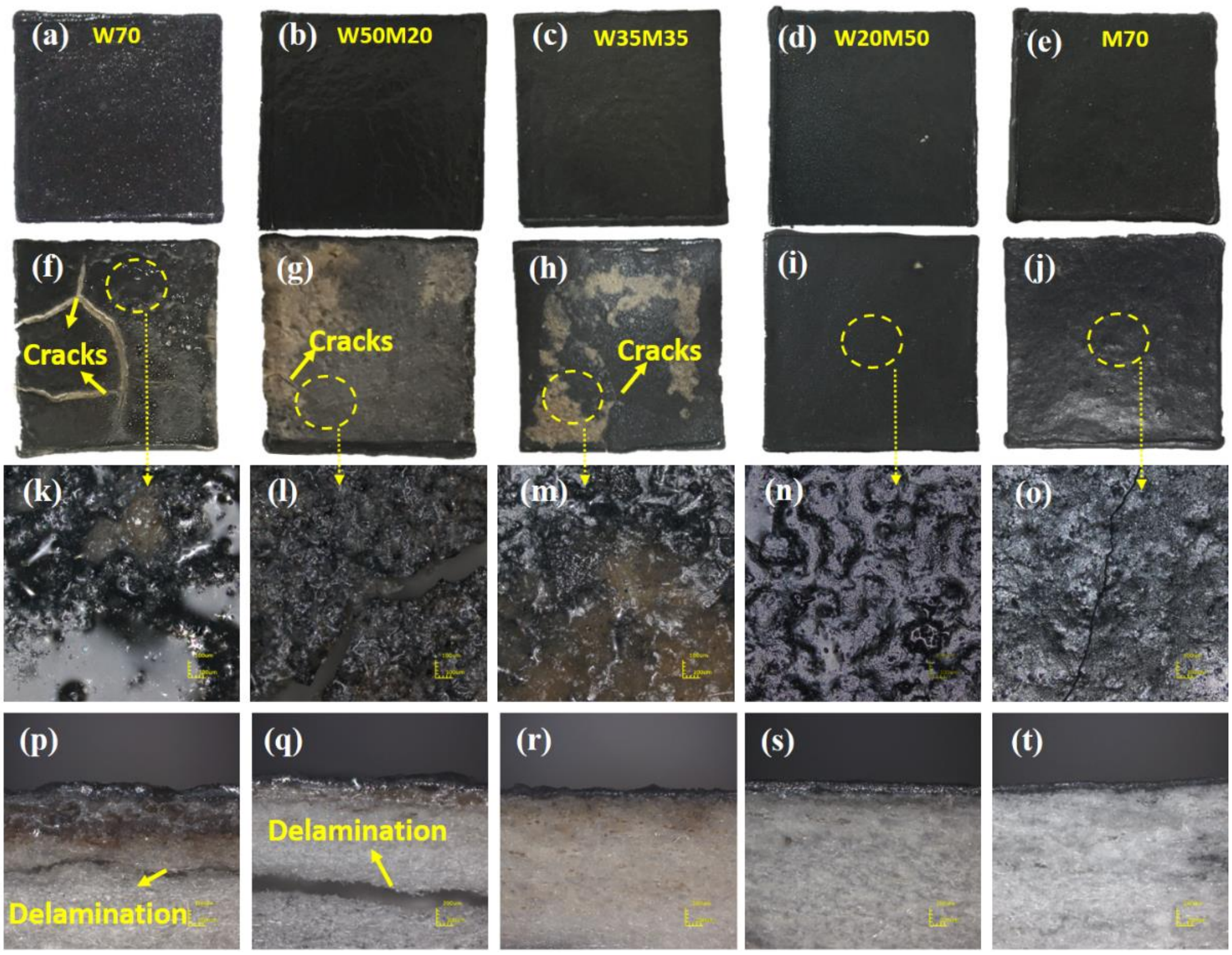

**Fig.12 Macrographs of the as–prepared coatings (a–e) and the coatings after oxidation at 1773 K for 50 h (f–j); The surface (k–o) and cross–section (p–t) CLSM images of the coatings after oxidation at 1773 K for 50 h**

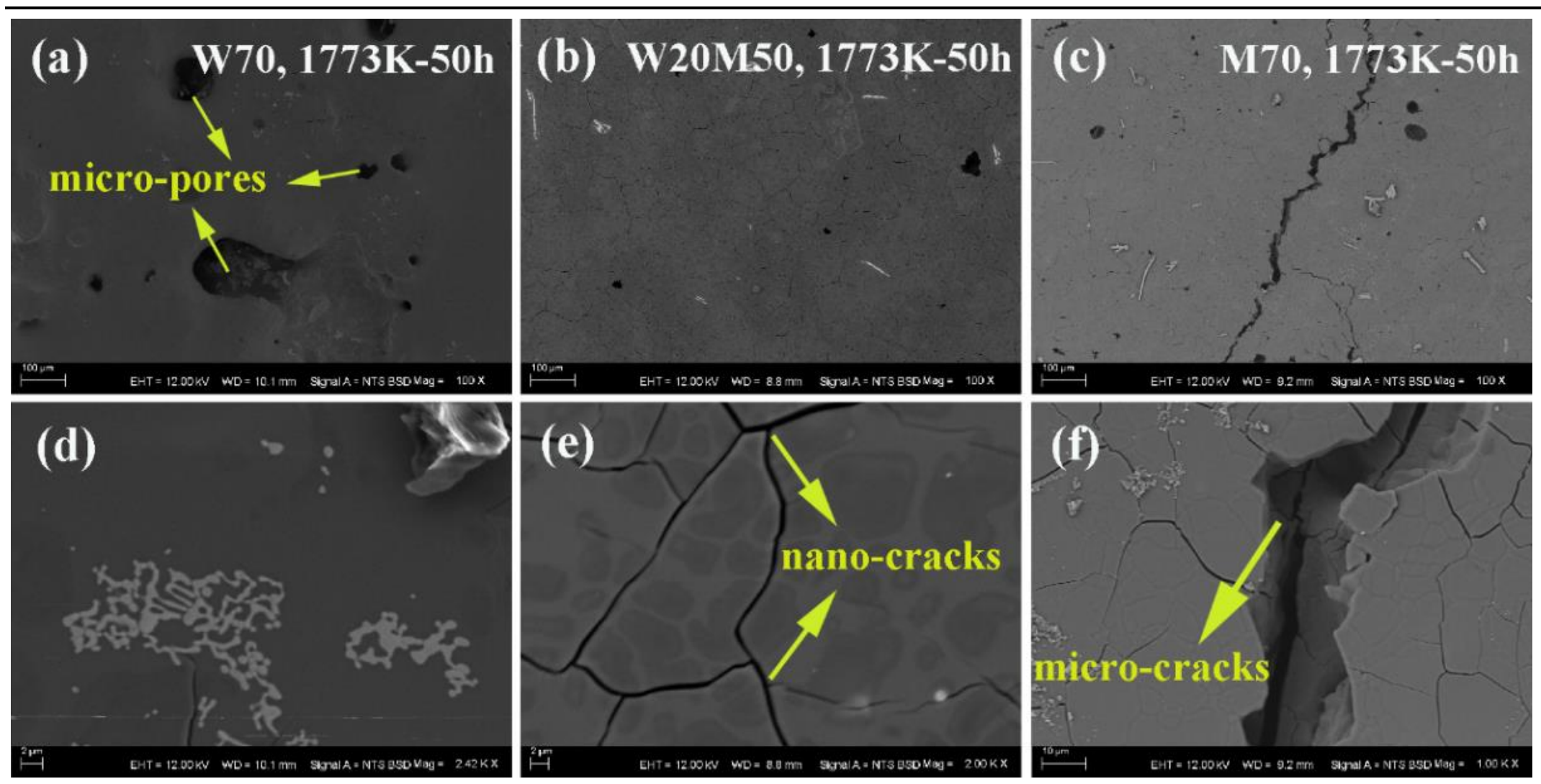

**Fig.13 SEM images of coatings after oxidation at 1773 K for 50 h in air: W70 (a,d); W20M50 (b,e); M70(c,f)**

Similarly, in M70 a deep micro-crack and some nano-sized cracks are evident (Fig. 13(f)) and are attributed to the high viscosity of the top $SiO_2$-rich glass layer which limits its ability to seal cracks. The BSE image of Fig. 14 shows the polished cross–section morphology and the corresponding elemental mapping of the W20M50 coating after oxidation. Though some surface nano-cracks are observed, no vertical cracks are evident for this composition.

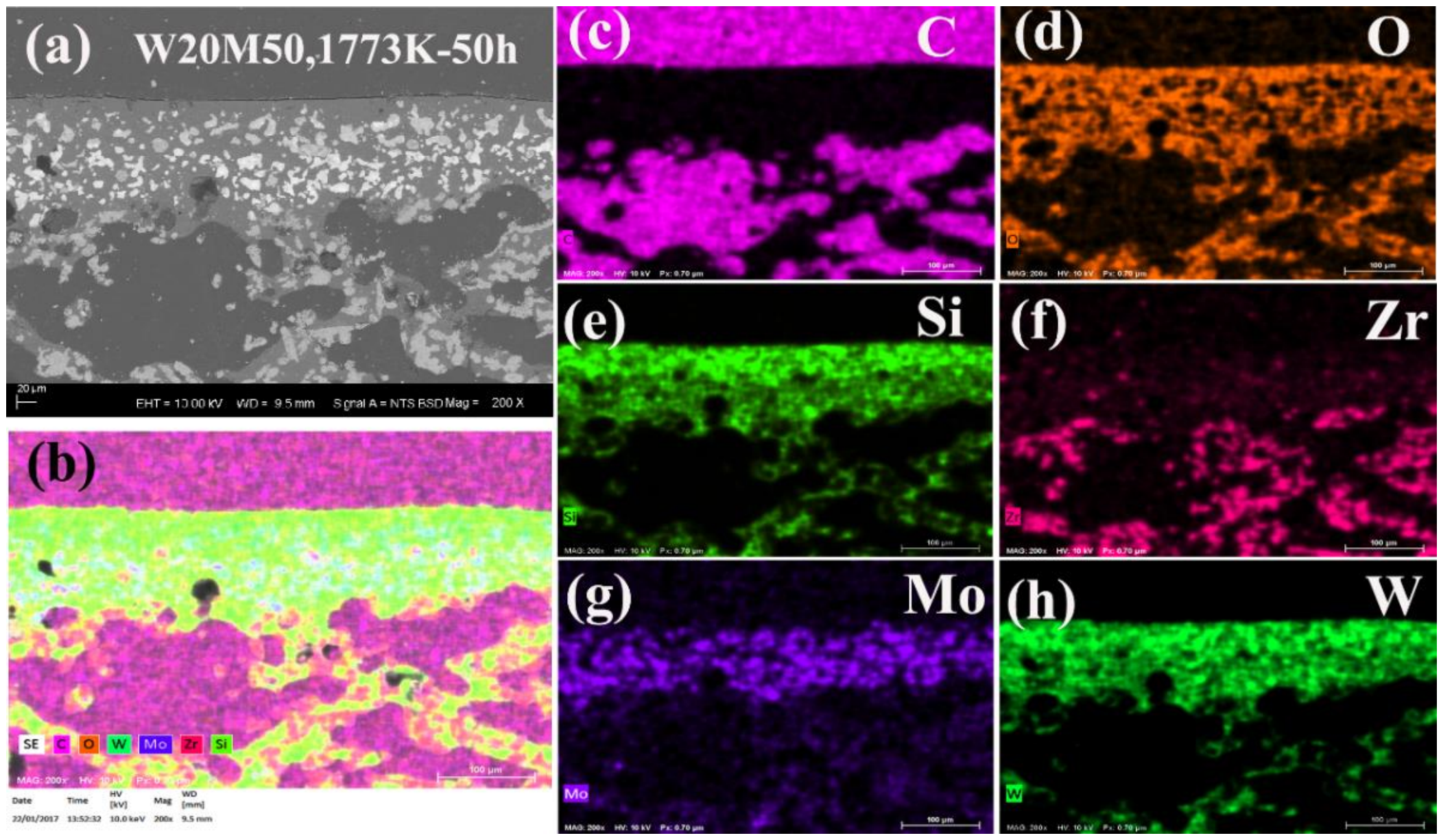

**Fig.14 (a) BSE image and (b–h) EDS mappings of the polished cross–section of W20M50 after oxidation at 1773 K for 50 h**

## 4. Discussion

### 4.1 Comparative performance

Towards applications in thermal protection systems, various silicide materials for oxidation resistant and high emissivity coatings have been investigated [33-36]. Systems studied thus far include YSZ-La-Mo-Si, $MoSi_2$-$WSi_2$-SiC-Si and $TaSi_2$-$MoSi_2$ with borosilicate glass. However, materials investigated to date have been limited in terms of their high temperature oxidation resistance. Following long-term high temperature treatment at temperatures similar to those employed in the present work, oxide phases, such as $SiO_2$ or $Ta_2O_5$, are ubiquitously observed to form on the surfaces of these materials, while diffraction peaks corresponding to the emissive agents are subsequently absent. The formation of oxidized phases and the consumption of the emissive phases limits the utility of such coatings for HEC applications.

In our previous work, $WSi_2$ was confirmed as a high emissivity material in theory by first principle calculations. Then a novel $WSi_2$-Si-glass coating using $WSi_2$ as an emissive agent was fabricated. However, after oxidation at 1773 K for 20 h, the emissivity dramatically declined and the coating was destroyed [14]. In order to improve the thermal resistance of the coating, $WSi_2$ and $MoSi_2$ were combined as double phase emissive agents in this work. Of the compositions studied, W20M50 exhibited the most promising performance. Diffraction patterns show that the emissive phases $WSi_2$ and $MoSi_2$ were still present in the surface of this coating following oxidation at 1773 K for 50 h, which is ascribed to the low oxygen diffusion rate of the $SiO_2$-based glass compound in the outmost layer and the self–healing ability of the inner layer of the coating over a wide temperature range. Based on the analysis of XRD, XPS and morphology results, it can be inferred that the thermal stability and appropriate viscosity of the top glass layer and the good self-healing ability of W20M50 help maintain the emissivity of the coating under a high temperature oxidizing atmosphere.

## 4.2 Oxidation mechanism

In general, the oxidation process consists of two stages: the initial fast oxidation stage, accompanied by significant defect formation and the second, slow stable oxidation stage, which is accompanied by defect healing. There are two sources of thermal stress in the coating during the oxidation process: (i) differential thermal expansion between the coating and the substrate (ii) aging stress between the newly formed surface oxide layer and inner coating layer. Thermal stress induced defects including cracks, spallation and delamination are observed to varying extents in all coatings, directly affecting the radiative property of the coatings after long–term high temperature oxidation. A further defect: micro pores, arise as the result of the volatilization of $B_2O_3$, $WO_3$ and $MoO_3$.

### 4.2.1 Defect formation

Firstly, the magnitude of the thermal stress arising from differential thermal expansion can be calculated by Eq. (13)[33]:

$$\delta = \frac{E \cdot \Delta\alpha \cdot \Delta T}{1 - v} \qquad (13)$$

in which $E$ is the elastic modulus , $\Delta\alpha$ is the difference between the coefficients of thermal expansion, $\Delta T$ refers to the temperature change and $v$ is Poisson's ratio. Differential thermal expansion between the fibrous $ZrO_2$ substrate ($\alpha$=10.2×10$^{-6}$/K[10]) and $WSi_2$ and $MoSi_2$ based coatings ($MoSi_2$: $\alpha$=9.32×10$^{-6}$/K[10]; $WSi_2$: $\alpha$=8.5× 10$^{-6}$/K[26]) is responsible for crack formation, particularly during the rapid cooling and heating stages. For the W70 coating, the larger $\Delta\alpha$ induces higher thermal stress relative to M70. Stress concentration at the interface with the substrate causes macro cracks and delamination, which are therefore observed more extensively in W70 (Fig. 12(f) and (p)). Secondly, according to the Pilling–Bedworth theory[37], the magnitude of the aging stress of the coating, which arises as the result of oxidative swelling, can be predicted using Pilling–Bedworth ratio (PBR), and the formula is:

$$\text{PBR} = \frac{V_{\text{oxide scale}}}{V_{\text{coating}}} \qquad (14)$$

where $V$ represents the molar volume. For the oxidation of $MoSi_2$, the calculated PBR values for the formation of Mo, $MoO_2$, $MoO_3$ and $Mo_5Si_3$, are 2.51, 2.94, 3.38 and 1.68, respectively [11]. Correspondingly, the calculated PBR values for the formation of W, $WO_3$ and $W_5Si_3$ products resulting from the oxidation of $WSi_2$ are 2.4, 3.27 and 1.96, respectively [14]. PBR＞1, corresponds to swelling, and compressive stress in the oxide scale, while tensile stress develops in scales with PBR＜1 [37]. Due to oxidative swelling, scales on $WSi_2$-$MoSi_2$-Si-glass coatings are in a state of compression irrespective of the oxidation route, and once sufficient coating stress is accumulated, cracks emerge on the coating surface. Accordingly, oxygen penetrates into the coatings through the aforementioned cracks and reacts with $WSi_2$, $W_5Si_3$, W, $MoSi_2$, $Mo_5Si_3$ and Mo according to reactions (7), (10)-(12) to form volatile $WO_3$ and $MoO_3$, which escape the coatings through these cracks leading to a continuous weight loss during the initial fast oxidation stage. To better understand the volatilization of $MoO_3$ and $WO_3$, the vapor pressures of the volatile oxides, which form as trimers $(MoO_3)_3$ and $(WO_3)_3$, were calculated by FactSage software and are plotted as a function of temperature (Fig. 15).

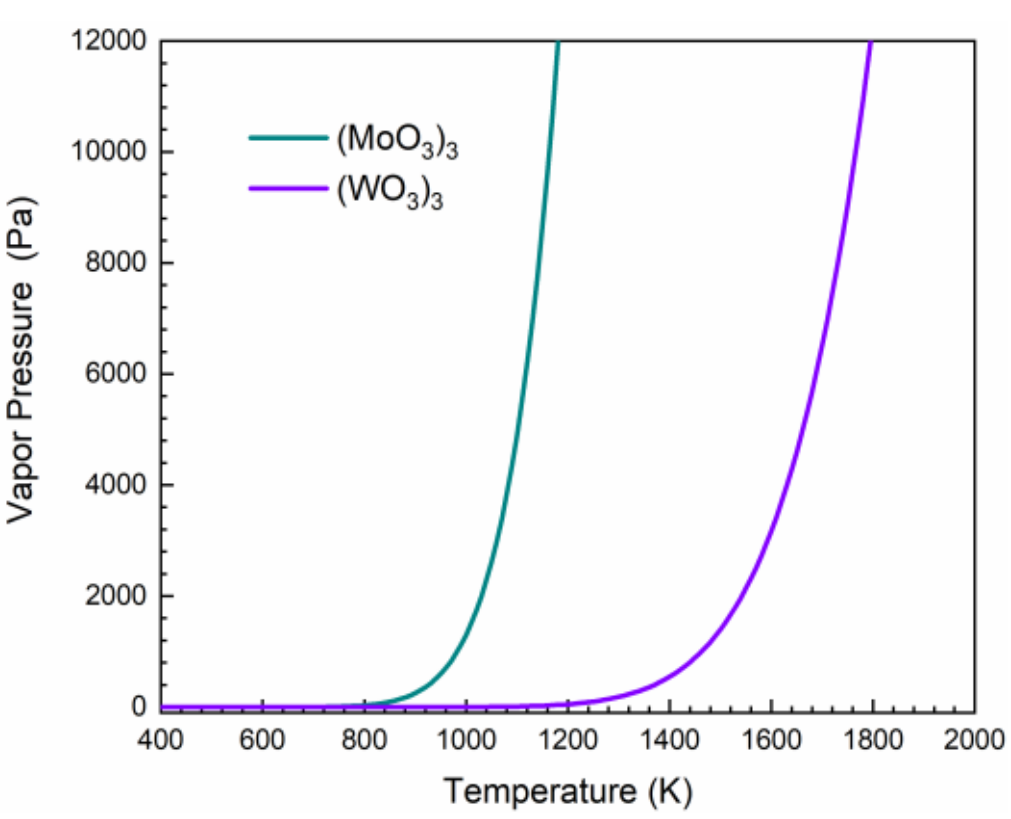

**Fig.15 Vapor pressure of $(MoO_3)_3$ and $(WO_3)_3$**

As the oxidation temperature increases, "pesting" initially starts at ~773K [38-40], with the volatilization of $MoO_3$ in the form of $(MoO_3)_3$. The evaporation of $MoO_3$ takes precedence over $WO_3$. In the intermediate temperature range (1073–1273 K), the evaporation and formation of $(MoO_3)_3$ occur simultaneously while $WO_3$ still remain on the surface due to the low vapor pressure[41]. As the temperature increases further (~ 1473 K), the vapor pressure of $(WO_3)_3$ increases, resulting in volatilization of both $(MoO_3)_3$ and $(WO_3)_3$.

### 4.2.2 Defect healing

Owing to the formation of flowable molten phases dispersed within the coatings at high temperatures, defects can be healed to a certain extent at various temperatures. $SiO_2$ and $B_2O_3$, the oxidation products of $SiB_6$ around 840–1240 K, will react to form a borosilicate glass, which can promote the densification of the coating during rapid sintering and seal the defects during long–term oxidation at relatively low temperatures. Another sintering additive, Si, can consume oxygen, thus further inhibiting the oxidation of $WSi_2$ and $MoSi_2$. Meanwhile, molten Si can seal micro cracks at high temperatures. The self-healing ability of coatings is determined not only by the amount of borosilicate glass but also its viscosity [42]. Yan et al. [43] correlated the viscosity of borosilicate glasses with temperature and boron content, as described in equation (15)

$$\log\eta = 3.11 - 19.2\exp(-24x) + \frac{1.68\times10^4}{T} + \frac{4.56\times10^4\exp(-22x)}{T}$$

(15)

. Viscosity is plotted as a function of temperature in Fig.16. Viscosity decreases with increasing boron content, especially at levels under 10 mol % $B_2O_3$. As the temperature increases to 1673 K, $B_2O_3$ evaporates and escapes from the borosilicate glass, leading to an increased viscosity of the surface glass layer. Compared with the rapid evaporation of $MoO_3$, the relatively slower evaporation of $WO_3$ is expected to allow sufficient time for borosilicate glass scale to flow and seal the surface[44]Based on the XPS analysis of O1s, the NBO concentration of W70 is the highest among W70, W20M50 and M70 coatings (Fig. 11(a)), implying lower viscosity.

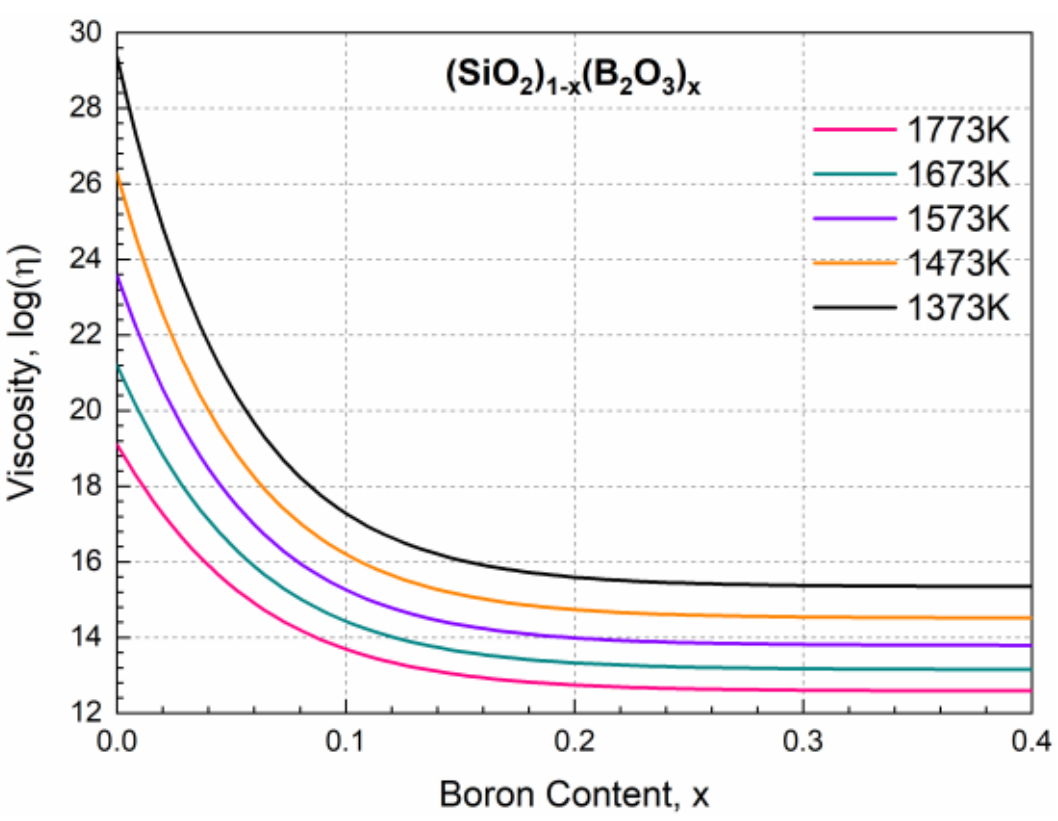

**Fig.16 The viscosity of borosilicate as a function of boron content**

The borosilicate glass in this material flows more readily and can effectively seal cracks (Fig. 13(a) and (d)). However, this material exhibits the lowest coefficient of thermal expansion as the result of lower glass phase viscosity and high $WSi_2$ content, with the resultant high thermal stresses inducing macroscopic cracks and delamination (Fig. 12(f) and (p)). In contrast, M70, which does not contains W phases, exhibits lower thermal stresses as the result of higher thermal expansion, and is more compatible with the substrate material. In M70, with the evaporation of $MoO_3$ and $B_2O_3$, increasing amounts of cristobalite are formed on coating surfaces as confirmed by XRD results (Fig.8), and the NBO concentration in M70 decreases to 7.47% after oxidation for 50h (Fig.11(c)). The lower fluidity of borosilicate glass with higher $SiO_2$ content limits effective crack sealing in this material (Fig. 12(o) and Fig. 13(c)), resulting in micro–sized cracks (Fig. 13(f)). Achieving optimal self-healing functionality necessitates a balance between glass phase viscosity at high temperatures and reduced thermal stresses at the coating interface. Combining $WSi_2$ and $MoSi_2$ as double phase emissive agents yields suitable viscosity for crack healing while maintaining acceptable levels of differential thermal expansion, thereby improving thermal oxidation resistance of coatings. Consequently, in the W20M50 material only nanoscale cracks were observed on the coating surface (Fig. 13(e)).

### 4.3 Summarized advantages of combined tungsten and molybdenum disilicides

Fig. 17 shows a brief schematic of the oxidation processes of the $WSi_2$–$MoSi_2$ based coatings according to the above analysis. In the initial fast oxidation stage, cracks are formed on the surfaces of coatings as the result of differential thermal expansion and aging stress. The volatilization of oxides like $MoO_3$, $WO_3$ and $B_2O_3$ leads to rapid mass loss. In parallel, oxidation reactions result in the formation of a passivating glass scale at the coating surface, which plays an important role in limiting further oxidation induced damage during the slow stable oxidation stage. The different viscosities of the surface glass layers result in different failure behaviors of the coatings.

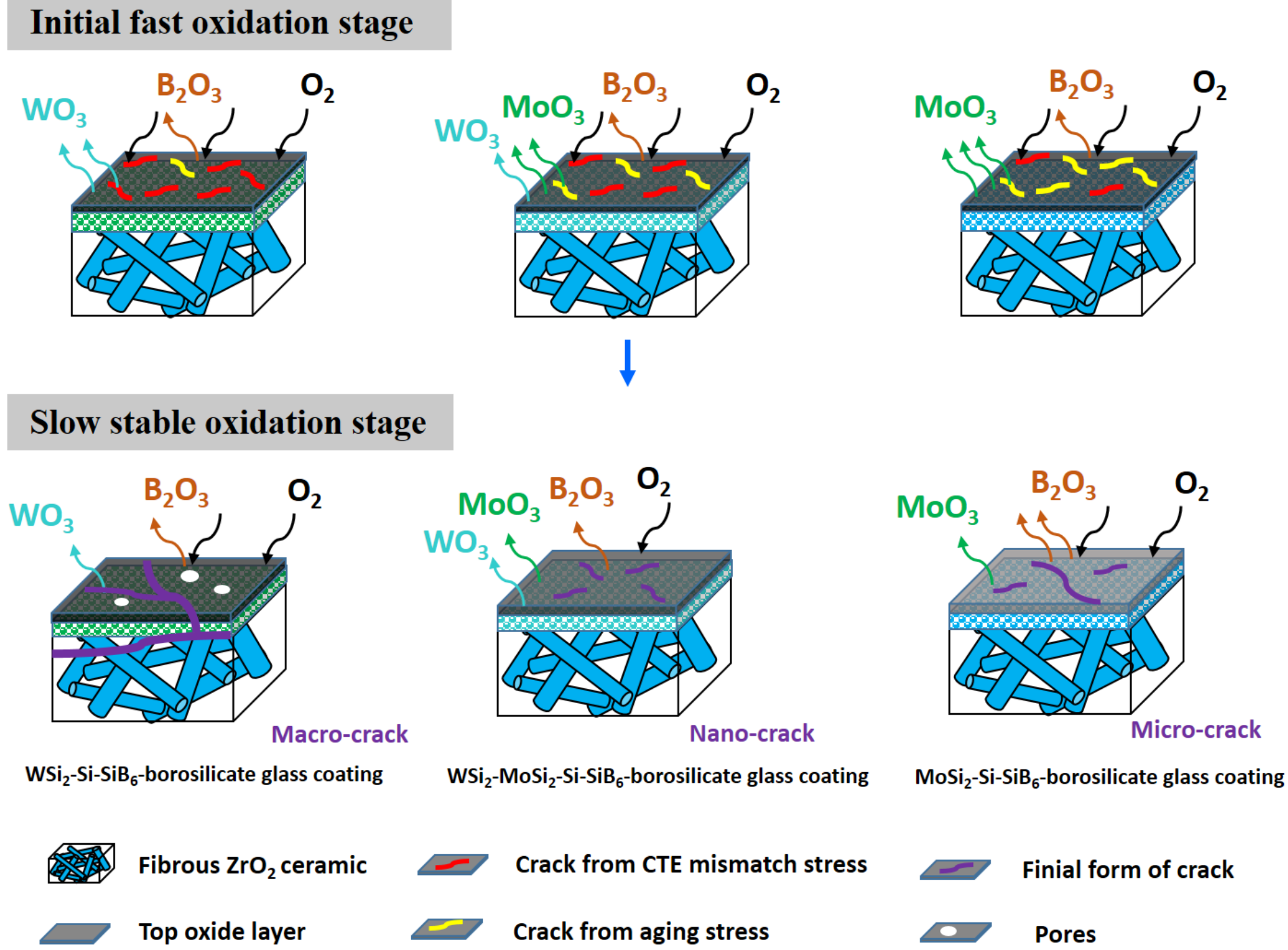

**Fig.17 Schematic oxidation processes of coated materials**

Although the glass layer in W70 exhibits a low viscosity and can more effectively seal mirco-cracks, the larger thermal stresses between the coating and the substrate leads to the formation of macro-cracks and delamination. In contrast, the M70 coating, with higher viscosity and worse fluidity cannot sufficiently seal the cracks, resulting in micro-cracks. Relative to coatings with only a single silicide phase, the advantages of double phase metal disilicides (combined $WSi_2$ and $MoSi_2$) include (1) better thermal expansion compatibility between the coating and the fibrous $ZrO_2$ substrate, and consequently less defect formation; (2) improved self-healing ability as the result of the oxidation product $WO_3$ limiting the evaporation of $MoO_3$, providing sufficient time for borosilicate glass scale flow, and the appropriate viscosity of the surface glass layer. Accordingly, coatings with both $WSi_2$ and $MoSi_2$ phases exhibited improved oxidation resistance and thermally robust emissivity under a high temperature oxidizing atmosphere, with highest performance levels found in W20M50 materials in the present work. Further compositional optimization and thermal resistance at higher temperatures and dynamic conditions where ablation is significant, remain ripe for further investigation.

## 5. Conclusion

(1) High emissivity multiphase coatings based on different compositions of $WSi_2$, $MoSi_2$, Si, $SiB_6$ and borosilicate glass have been prepared on fibrous $ZrO_2$ by slurry dipping and a subsequent high temperature rapid sintering method. $MoSi_2$ and $WSi_2$ particles are dispersed in the continuous borosilicate glass in the coatings. The presence of a glass layer limits oxygen diffusion and results in the formation of thermally stable W, Mo, $W_5Si_3$ and $Mo_5Si_3$ phases from the oxidation of disilicide phases on the surface of the coating.

(2) Of the compositions studies W20M50, comprising 20 wt% $WSi_2$ and 50 wt% $MoSi_2$, presents the highest oxidation resistance, with only 10.06 mg/cm$^2$ mass loss and 4.0 % radiative loss in the wavelength regime 1.27-1.73 μm after oxidation at 1773 K for 50 h. In this material, defect formation was observed to occur during the initial fast oxidation stage, while defect healing occurred during the slow stable oxidation.

(3) The advantages of double phase metal silicides ($WSi_2$ and $MoSi_2$) include (i) better thermal compatibility with the substrate and (ii) appropriate viscosity of the surface glass layer combined with sufficient defect sealing time.

**Acknowledgment**

This work was financially supported by the Program for Changjiang Scholars and Innovative Research Team in University (No. IRT_15R35), the National Natural Science Foundation of China (No.51602151), the Natural Science Foundation of Jiangsu Province (No.BK20161003), the Major Program of Natural Science Fund in Colleges and Universities of Jiangsu Province (No.15KJA430005), Priority Academic Program Development of Jiangsu Higher Education Institutions (PAPD), China Scholarship Council (CSC, 201608320159).